\documentclass[12pt, a4paper]{article}

\usepackage[margin=1in]{geometry}
\usepackage{amsmath, amssymb}
\usepackage{graphicx}
\usepackage{booktabs}
\usepackage{array}
\usepackage{multirow}
\usepackage[hyphens]{url}
\usepackage{hyperref}
\usepackage{cite}
\usepackage{caption}
\usepackage{subcaption}
\usepackage{float}
\usepackage{enumitem}
\usepackage{tikz}
\usetikzlibrary{shapes.geometric, arrows.meta, positioning, fit, backgrounds}
\usepackage{microtype}
\usepackage{xcolor}

\hypersetup{colorlinks=true, citecolor=blue, linkcolor=black, urlcolor=blue}

\newif\ifdraft
\draftfalse
\newcommand{\needsupdate}[1]{\ifdraft\textcolor{red}{#1}\else#1\fi}

\tikzset{
  block/.style={rectangle, draw=black, text width=4.0cm, align=center,
                minimum height=0.9cm, font=\small},
  blockdash/.style={rectangle, draw=black, dashed, text width=4.0cm, align=center,
                minimum height=0.9cm, font=\small},
  gate/.style={rectangle, draw=black, text width=2.0cm, align=center,
               minimum height=0.7cm, font=\small},
  dec/.style={diamond, draw=black, aspect=2, align=center, font=\small,
              inner sep=1pt},
  term/.style={rectangle, draw=black, rounded corners, text width=2.6cm,
               align=center, minimum height=0.7cm, font=\small},
  arr/.style={-{Stealth[length=2mm]}, thick},
}

\begin{document}

% -- Title ---------------------------------------------------
% NOTE ON STRUCTURE: each block below is its own center environment
% with the font size applied as a DECLARATION, not inside braces.
% \end{center} supplies the \par while the size is still in effect,
% so line spacing is correct without closing a braced group with an
% explicit \par. That construct, combined with \\[dim], fails on
% TeX Live 2025 with "Missing } inserted". Do not reintroduce it.
\begin{center}
  \Large\bfseries
  A Hybrid Two-Stage Machine Learning Pipeline for\\[5pt]
  Fault Detection and Classification in Power Transmission Systems
\end{center}

\vspace{-4pt}

\begin{center}
  \large
  Sahil Manikshete$^{1}$, Atharva Gujarathi$^{1}$, Thanh Long Vu$^{2}$,\\[4pt]
  Akhtar Hussain$^{3}$, and Van-Hai Bui$^{4}$
\end{center}

\vspace{-4pt}

\begin{center}
  \small
  $^{1}$Department of Computer and Information Science,
  University of Michigan-Dearborn, Dearborn, MI, USA\\
  $^{2}$Pacific Northwest National Laboratory, Richland, WA, USA\\
  $^{3}$Department of Electrical and Computer Engineering,
  Universit\'e Laval, Qu\'ebec City, QC, Canada\\
  $^{4}$Department of Electrical and Computer Engineering,
  University of Michigan-Dearborn, Dearborn, MI, USA\\[4pt]
 
\end{center}

\vspace{8pt}\hrule\vspace{10pt}

% -- Abstract (244 words; IEEE limit 250) ---------------------
\noindent\textbf{Abstract}---%
Rapid and accurate fault detection in high-voltage transmission networks is essential for grid
reliability and equipment protection. Transmission fault datasets are frequently imbalanced, and
certain fault types produce electrical signatures that fall within the normal operating envelope,
causing single-model classifiers to fail on safety-critical cases. This paper proposes a hybrid
two-stage machine learning pipeline that decouples detection from classification. Stage~1 combines
an Isolation Forest anomaly detector with an optional supervised binary detector through an
OR-fusion rule; the supervised branch is allocated automatically during training for any fault class
the anomaly detector cannot resolve, and is omitted when no such class exists. Stage~2 applies a
Random Forest multiclass classifier only to samples flagged by Stage~1. Feature engineering is
expressed as a per-measurement-point operator mapping six raw channels to eighteen features,
including zero-sequence symmetrical components derived from Fortescue's theorem, yielding $18L$
features for $L$ measurement points. On the TLFaultDataset ($L = 7$, 126 features, 578{,}923
samples) the pipeline raises Line-fault end-to-end accuracy from 31.3\% to \needsupdate{95.8\%}. On
an independent single-point dataset ($L = 1$, 18 features, 12{,}001 samples) the same framework
attains 97.25\% end-to-end accuracy across all classes including normal operation, exceeding the
TLFed federated benchmark of 94.84\% without GPU or federated infrastructure, at 0.05~ms per sample
on CPU. Ablation on both datasets shows zero-sequence features resolving the three-phase versus
three-phase-to-ground ambiguity, raising the F1-score of that class pair from 0.39 to 0.997. The
direction of the zero-sequence signature is found to be system-dependent, motivating a learned
decision boundary in place of a fixed relay threshold.


\vspace{8pt}\hrule\vspace{10pt}

% ============================================================
\section{Introduction}
% ============================================================

Power transmission systems form the backbone of modern electrical infrastructure, delivering
electricity over long distances from generation sources to distribution networks. Faults caused by
lightning strikes, equipment failure, conductor contact, or insulation breakdown can trigger
cascading failures, equipment damage, and widespread outages with significant economic and social
consequences~\cite{blackburn2006}. Rapid and accurate fault detection and classification is
therefore essential for minimizing downtime, enabling targeted corrective action, and maintaining
grid reliability. Traditional protection methods such as distance relays, differential protection, and
overcurrent relays rely on fixed thresholds and simplified assumptions about system
behavior~\cite{gers2011}. While effective in conventional systems, these approaches face growing
challenges in modern networks characterized by distributed energy resources, bidirectional power
flow, inverter-based generation, and dynamically changing fault current
levels~\cite{mohammadi2024}.

The core challenge is twofold. First, transmission fault datasets frequently exhibit severe class
imbalance: in the TLFaultDataset used as the primary evaluation system in this study, 88.1\% of
samples represent normal operation while the rarest fault type accounts for under 1\% of all samples.
Classifiers trained on such data become biased toward the majority class and fail to detect rare but
critical faults. Second, certain fault types---particularly transmission Line faults---produce phase
voltage and current signatures that are nearly indistinguishable from normal operation. A
single-model approach achieves only 31.3\% end-to-end detection accuracy for Line faults, which is
insufficient for protection use.

Machine learning has been widely explored as a data-driven complement to traditional protection, and
the literature spans classical supervised classifiers, signal-transform front ends, ensemble methods,
and, more recently, deep and federated architectures. Supervised methods including Random Forest
(RF), Support Vector Machine (SVM), and $k$-Nearest Neighbors (KNN) have been applied to fault
classification, typically achieving high accuracy on balanced or Synthetic Minority Over-sampling
Technique (SMOTE) augmented data~\cite{chen2021, hu2020}. These methods perform well when the class
distribution is controlled, but their reported accuracies are usually obtained on rebalanced data and
do not directly reflect performance under the severe native imbalance of transmission fault records.

A second line of work improves the input representation rather than the classifier. Feature
extraction based on the Discrete Wavelet Transform (DWT) coupled with an Artificial Neural Network
(ANN) has been used for fast transmission-line fault detection~\cite{abdullah2017}, and
time--frequency descriptors derived from the Hilbert--Huang Transform have been combined with a
Convolutional Neural Network (CNN) for distribution-system fault classification~\cite{guo2021}.
Ensemble methods that combine several base classifiers through voting and stacking have reported
accuracy approaching 99\% on controlled benchmarks~\cite{shukla2019}. Two-stage sequential
architectures, in which an unsupervised step precedes a supervised classifier, have also been
explored~\cite{biswal2021}, as have anomaly-detection approaches using the Isolation Forest for
power-quality monitoring and photovoltaic fault localization~\cite{ahmed2019}. These studies
establish that both detection and classification can be handled by learned models, but they generally
treat all fault classes as equally separable and do not isolate the classes whose signatures overlap
with normal operation. A further limitation shared across this literature is that methods are
typically reported on a single dataset, so it is rarely established whether an architecture or a
feature set transfers to a different network.

A directly relevant prior work is TLFed~\cite{custodio2023}, a federated learning system using a
one-dimensional Convolutional Neural Network with Long Short-Term Memory (1D-CNN-LSTM) architecture
across local client nodes with Federated Averaging (FedAvg) aggregation. TLFed uses the same
TLFaultDataset and seven fault classes, achieving 94.84\% classification accuracy and 92.50\%
location accuracy across 21 federated clients, motivated by data-privacy concerns in smart grids. It
provides the closest baseline for the present study because it shares the dataset and class
definitions, but it treats all seven classes uniformly and relies on GPU-trained deep models
distributed across clients.

Despite these advances, existing approaches share several limitations. Methods relying on SMOTE or
class weighting improve average accuracy but still struggle with fault types whose signatures overlap
with normal operation, since synthetic oversampling cannot capture the true distribution of rare
faults. Single-model architectures must simultaneously solve detection (fault versus no fault,
requiring maximum recall) and classification (which fault type, requiring per-class precision), two
objectives that are difficult to optimize jointly. The Isolation Forest, despite being designed for
anomaly detection, achieves only 0.9\% detection on Line faults because these do not appear anomalous
in the raw feature space. TLFed achieves strong overall accuracy but treats all seven classes
uniformly, with no mechanism to handle the structural blind spot of near-normal Line faults. Finally,
generic statistical features such as mean and standard deviation do not encode the physical mechanisms
that distinguish fault types; in particular, distinguishing three-phase faults with and without ground
involvement requires zero-sequence symmetrical components grounded in Fortescue's decomposition
theorem~\cite{fortescue1918}.

This paper proposes a hybrid two-stage machine learning pipeline that addresses these limitations.
The central premise is that different fault types require different detection strategies, and that
detection and classification are better solved as two sequential but independent problems. In
Stage~1, an Isolation Forest trained exclusively on normal data detects general faults as anomalies;
where a fault class remains invisible to this detector, a dedicated supervised binary classifier is
allocated for that class and fused by OR logic to maximize recall. In Stage~2, a Random Forest
multiclass classifier receives only the samples flagged by Stage~1 and determines the fault type. The
pipeline is supported by physics-informed feature engineering that augments the raw per-point
measurements with features capturing signal energy, phase asymmetry, and zero-sequence components.
The framework is evaluated on two independent datasets differing in network size, voltage scale,
simulation software, and class balance.

The principal contributions of this work are as follows.
\begin{enumerate}[leftmargin=*, itemsep=4pt]
  \item \textbf{A two-stage architecture with conditional supervised detection.} An unsupervised
        Isolation Forest is fused by OR logic with a dedicated supervised detector that is allocated
        automatically, during training, for any fault class the anomaly detector fails to resolve.
        On the TLFaultDataset this rule identifies the Line-fault class, raising its end-to-end
        accuracy from 31.3\% to \needsupdate{95.8\%}; on the second dataset it identifies a different
        class, indicating that the mechanism is not specific to one system.

  \item \textbf{A network-size-independent feature construction.} Feature engineering is expressed as
        a per-measurement-point operator mapping six raw channels to eighteen
        features root-mean-square (RMS) energy measures, phase imbalance ratios, and zero-sequence
        symmetrical components applied identically at each of $L$ measurement points to give $18L$
        features. The same construction is applied without modification at $L = 7$ and $L = 1$.

  \item \textbf{Zero-sequence components as the discriminator for ground involvement, with a learned
        rather than fixed boundary.} Zero-sequence features supply the information that separates
        three-phase from three-phase-to-ground faults, an ambiguity that raw phase measurements
        cannot resolve; ablation raises the F1-score of this class pair from 0.39 to 0.997 on the
        second dataset. The magnitude and direction of the zero-sequence signature are shown to be
        system-dependent, since the residual current depends on fault balance and grounding impedance
        as well as on the presence of a ground path. A learned decision boundary therefore
        generalizes across systems where a fixed threshold does not.
\end{enumerate}

% ============================================================
\section{System Model}
% ============================================================

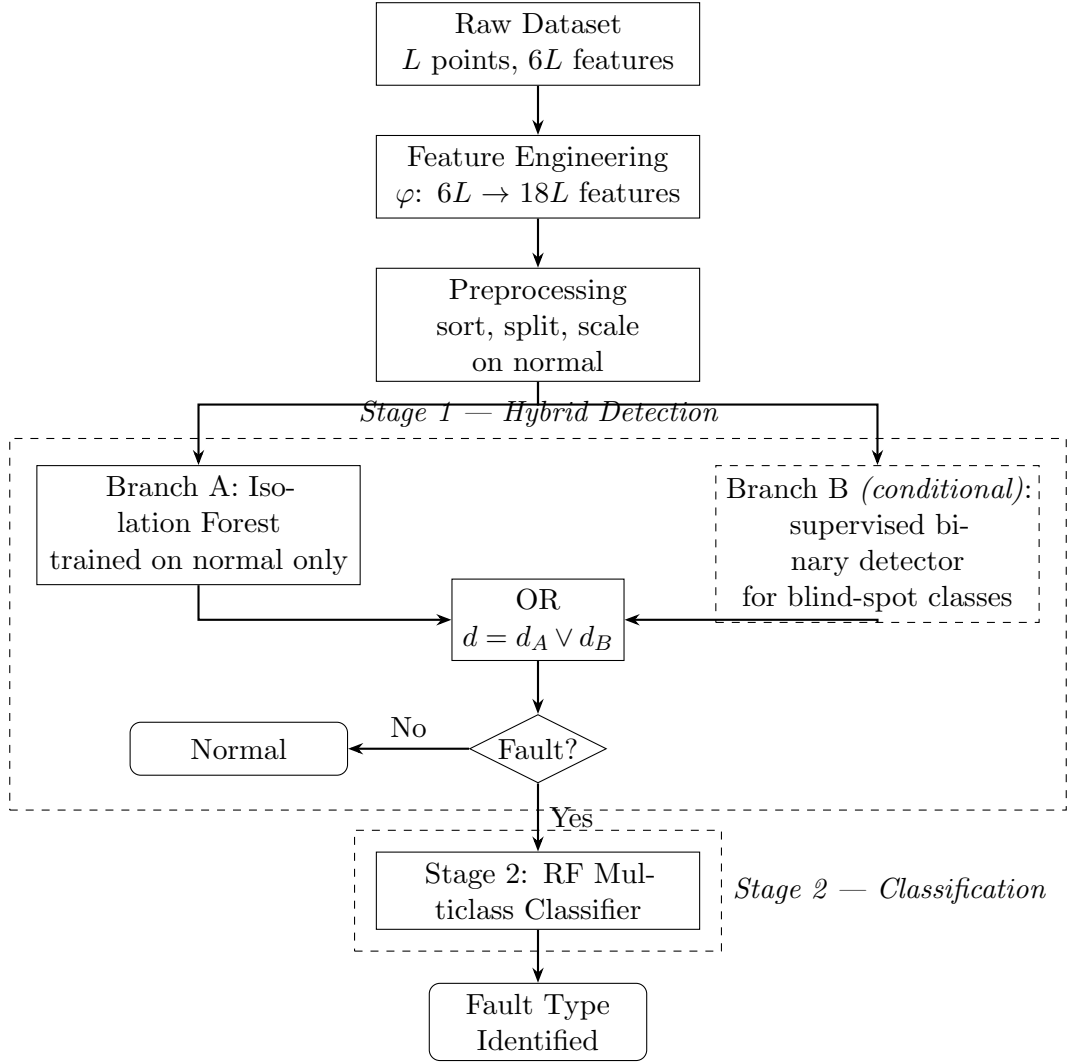
\begin{figure}[H]
\centering
\begin{tikzpicture}[node distance=0.65cm and 1.0cm]

  \node[block] (input) {Raw Dataset\\$L$ points, $6L$ features};
  \node[block, below=of input] (feat) {Feature Engineering\\$\varphi$: $6L \rightarrow 18L$ features};
  \node[block, below=of feat] (prep) {Preprocessing\\sort, split, scale on normal};

  \node[block, below left=1.1cm and 0.2cm of prep] (iso)
        {Branch A: Isolation Forest\\trained on normal only};
  \node[blockdash, below right=1.1cm and 0.2cm of prep] (rf)
        {Branch B \emph{(conditional)}:\\supervised binary detector\\for blind-spot classes};

  \node[gate, below=2.6cm of prep] (or) {OR\\$d = d_A \vee d_B$};
  \node[dec, below=0.7cm of or] (q) {Fault?};
  \node[term, left=1.6cm of q] (normal) {Normal};

  \node[block, below=0.9cm of q] (s2)
        {Stage 2: RF Multiclass Classifier};
  \node[term, below=0.7cm of s2] (out) {Fault Type Identified};

  \draw[arr] (input) -- (feat);
  \draw[arr] (feat) -- (prep);
  \draw[arr] (prep.south) -- ++(0,-0.3) -| (iso.north);
  \draw[arr] (prep.south) -- ++(0,-0.3) -| (rf.north);
  \draw[arr] (iso.south) |- (or.west);
  \draw[arr] (rf.south) |- (or.east);
  \draw[arr] (or) -- (q);
  \draw[arr] (q) -- node[above]{\small No} (normal);
  \draw[arr] (q) -- node[right]{\small Yes} (s2);
  \draw[arr] (s2) -- (out);

  \begin{scope}[on background layer]
    \node[draw=black, dashed, fit=(iso)(rf)(or)(q), inner sep=10pt,
          label={[font=\small\itshape]above:Stage 1 --- Hybrid Detection}] {};
    \node[draw=black, dashed, fit=(s2), inner sep=8pt,
          label={[font=\small\itshape]right:Stage 2 --- Classification}] {};
  \end{scope}

\end{tikzpicture}
\caption{Two-stage fault detection and classification pipeline. Branch~B is drawn with a dashed
         outline to indicate that it is conditional: it is instantiated only for fault classes the
         Isolation Forest fails to detect, as determined by the allocation rule of
         Section~\ref{sec:branchb}.}
\label{fig:pipeline}
\end{figure}

The proposed framework is a hybrid two-stage machine learning pipeline designed around the principle
that fault detection and fault-type identification are distinct problems that benefit from independent
optimization. The method is formulated generally for any transmission system that exposes three-phase
current and voltage measurements at one or more points: given a network with $L$ measurement points,
each providing three-phase current and voltage readings, the pipeline applies the same feature
construction, detection, and classification stages irrespective of the specific network topology.
After feature engineering and preprocessing, the enriched data passes through two sequential stages:
Stage~1 determines whether a fault is present, and Stage~2 identifies the fault type for all samples
flagged by Stage~1. Fig.~\ref{fig:pipeline} illustrates the complete architecture.

\subsection{Stage 1: Hybrid Fault Detection}

Stage~1 answers the binary question of whether a fault is present, prioritizing recall over
precision: it is designed to capture every fault even at the cost of some false alarms, since a
missed fault carries substantially greater consequences than an unnecessary investigation.

\textbf{Branch A --- Isolation Forest.} The Isolation Forest~\cite{liu2008} is an unsupervised
ensemble that builds random trees, isolating each sample by randomly selecting a feature and split
point. Anomalous samples require fewer splits because they are sparse and differ from the majority.
The anomaly score is inversely related to the average path length across trees:
\begin{equation}
  s(x, n) = 2^{-\dfrac{E[h(x)]}{c(n)}},
  \qquad
  c(n) = 2H(n-1) - \dfrac{2(n-1)}{n},
\end{equation}
where $h(x)$ is the path length required to isolate $x$, $c(n)$ normalizes by the average path length
of an unsuccessful binary-search-tree query, and $H(\cdot)$ is the harmonic number. The detector is
trained exclusively on normal samples, so any measurement deviating sufficiently from the learned
normal distribution is flagged. The standardization transform is fitted only on normal training
samples; including fault data would shift the baseline and reduce sensitivity to anomalies.

The decision threshold applied to $s(x,n)$ is selected on a validation split held out from the
training data, at the operating point maximizing precision subject to a recall floor, and is then
frozen. The test set is used only once, for final evaluation.

\subsection{Branch B: Conditional Supervised Detection}
\label{sec:branchb}

Certain fault classes produce measurements that lie inside the normal operating envelope and are
therefore invisible to an anomaly detector trained on normal data, irrespective of its capacity or
contamination setting. Rather than assuming in advance which class this will be, the framework
determines it empirically.

Let $\mathcal{F}$ denote the set of fault classes and let $r_c$ be the Branch~A detection rate for
class $c$, measured on the validation split at the frozen operating point. The set of blind-spot
classes is
\begin{equation}
  \mathcal{B} = \{\, c \in \mathcal{F} \;:\; r_c < \tau \,\},
  \label{eq:allocation}
\end{equation}
for an allocation threshold $\tau$. For each $c \in \mathcal{B}$ a dedicated binary Random
Forest~\cite{breiman2001} detector is trained to separate class $c$ from all other samples, using all
available training samples of class $c$ together with a bounded random sample of the remainder and
balanced class weighting. The Random Forest aggregates votes from $B$ trees:
\begin{equation}
  \hat{p}_c(x) = \frac{1}{B}\sum_{b=1}^{B} \mathbb{1}\!\left[T_b(x) = c\right].
\end{equation}

If $\mathcal{B} = \emptyset$, no supervised detector is instantiated and Stage~1 reduces to Branch~A
alone. Branch~B is therefore an optional refinement of the two-stage architecture rather than a
structural requirement of it.

\textbf{OR-fusion rule.} Let $d_A(x) \in \{0,1\}$ denote the Branch~A decision and $d_{B,c}(x)$ the
decision of the detector allocated to class $c$. The fused Stage~1 decision is
\begin{equation}
  d_{\text{S1}}(x) = d_A(x) \;\vee\; \bigvee_{c \in \mathcal{B}} d_{B,c}(x).
  \label{eq:orfusion}
\end{equation}
A sample is flagged as a fault if any branch raises an alarm, so that no single branch's blind spot
determines the overall outcome. Both branches are evaluated for every sample at inference; the
conditionality in~\eqref{eq:allocation} applies during training only.

\subsection{Stage 2: Fault Type Classification}

Stage~2 receives only samples flagged as faults by Stage~1 and performs multiclass classification
over the fault classes present in the dataset. A Random Forest classifier is trained exclusively on
true fault samples in the training set and never processes normal data. Let $\mathcal{C}$ denote the
set of fault classes. Each of the $B$ trees casts a vote, and the predicted class is the majority
vote across the ensemble,
\begin{equation}
  \hat{y}(x) = \operatorname*{arg\,max}_{c \in \mathcal{C}}
               \sum_{b=1}^{B} \mathbb{1}\!\left[T_b(x) = c\right],
  \label{eq:stage2}
\end{equation}
where $T_b(x)$ is the class predicted by tree $b$. This focused training allows the classifier to
specialize in distinguishing fault types rather than separating faults from normal operation. The
separation provides several advantages: each stage can be independently evaluated, tuned, and
replaced; Stage~2 operates on a cleaner input space; the recall-precision trade-off at Stage~1 does
not affect Stage~2 quality; and the design mirrors deployed protection systems, in which detection
and identification are handled by different relay functions.

\subsection{Feature Engineering}
\label{sec:feateng}

Feature engineering is applied before Stage~1 and is defined as an operator $\varphi$ acting on a
single measurement point. Given the three-phase currents $(I_a, I_b, I_c)$ and voltages
$(V_a, V_b, V_c)$ at one point, $\varphi$ produces eighteen features: the six raw channels, eight
statistical descriptors, and four zero-sequence quantities. For a network exposing $L$ measurement
points, $\varphi$ is applied at each point and the results are concatenated, giving
\begin{equation}
  \dim\!\left(\varphi^{\otimes L}\right) = 18L
  \label{eq:featcount}
\end{equation}
features in total. The construction is therefore independent of network size: no component of it
refers to a particular topology or number of lines.

\textbf{Statistical features.} At each point, four statistics are computed from the three-phase
currents and the same four from the three-phase voltages, giving eight statistical features per
point. The RMS value summarizes signal energy,
\begin{equation}
  \mathrm{RMS} = \sqrt{\frac{I_a^2 + I_b^2 + I_c^2}{3}},
  \label{eq:rms}
\end{equation}
the maximum and mean absolute values capture peak and average magnitude,
\begin{align}
  \text{Max abs}  &= \max\!\left(|I_a|, |I_b|, |I_c|\right), \label{eq:maxabs}\\[3pt]
  \text{Mean abs} &= \tfrac{1}{3}\left(|I_a| + |I_b| + |I_c|\right), \label{eq:meanabs}
\end{align}
and the imbalance ratio quantifies phase asymmetry,
\begin{equation}
  \text{Imbalance ratio} =
    \frac{\max(|I_a|,|I_b|,|I_c|) - \min(|I_a|,|I_b|,|I_c|)}
         {\operatorname{mean}(|I_a|,|I_b|,|I_c|) + \epsilon},
  \label{eq:imbalance}
\end{equation}
where $\epsilon$ is a small constant preventing division by zero. The imbalance ratio is the most
informative statistical feature for Stage~1 anomaly detection: during a fault the affected phases
behave differently from unaffected phases, and the ratio magnifies this asymmetry, rendering
anomalies more visible to the Isolation Forest.

\textbf{Zero-sequence features.} At each point, the signed and absolute zero-sequence current and
voltage are computed from the symmetrical-component (Fortescue) decomposition,
\begin{equation}
  I_0 = \frac{I_a + I_b + I_c}{3},
  \qquad
  V_0 = \frac{V_a + V_b + V_c}{3}.
  \label{eq:zeroseq}
\end{equation}
Recording the signed and absolute values of both $I_0$ and $V_0$ gives four zero-sequence features
per point. Zero-sequence current is the residual that flows when the three phase currents do not sum
to zero, which requires both a path to ground and an asymmetry among the phase currents. It therefore
carries information about ground involvement that raw phase measurements do not expose directly; the
quantitative behavior of this signature across the two evaluation systems is examined in
Section~\ref{sec:zerosysdep}.

\textbf{Feature count.} Each measurement point contributes six raw channels, eight statistical
descriptors, and four zero-sequence quantities, giving $6 + 8 + 4 = 18$ per point and $18L$ in total
by~\eqref{eq:featcount}. For the TLFaultDataset, $L = 7$ gives $42 + 56 + 28 = 126$ features; for the
second dataset, $L = 1$ gives 18. The count is a direct consequence of network size rather than an
arbitrary design choice. Each engineered feature encodes a specific physical mechanism---signal
energy, peak magnitude, phase asymmetry, or ground-path involvement---so the expanded representation
adds discriminative physical information without introducing redundant or purely synthetic
dimensions. A feature may nonetheless be uninformative on a particular system: on the second dataset
the phase voltages sum identically to zero, so $V_0$ and $|V_0|$ are constant and 16 of the 18
features carry information. This condition is detected automatically by a scale-relative variance
check and is reported rather than concealed.

\begin{figure}[H]
  \centering
  \includegraphics[width=0.9\linewidth]{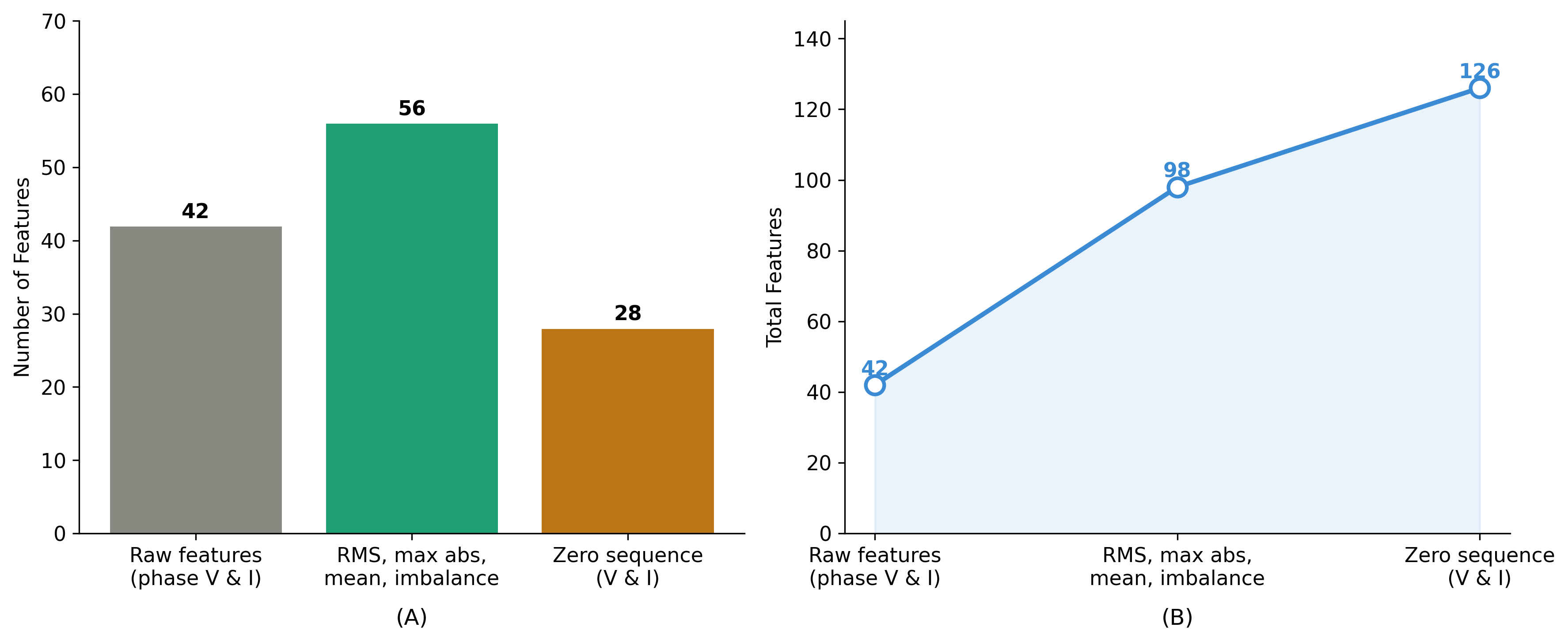}
  \caption{Feature engineering pipeline for $L = 7$: (a) number of features added at each step;
           (b) cumulative feature count, growing from 42 raw measurements to 126 in total.}
  \label{fig:feat}
\end{figure}

\subsection{Evaluation Metrics}
\label{sec:metrics}

Performance is reported using standard classification metrics derived from the counts of true
positives ($TP$), true negatives ($TN$), false positives ($FP$), and false negatives ($FN$). For
Stage~1 detection, the fault class is treated as positive, so recall measures the fraction of true
faults captured and the false alarm rate measures the fraction of normal samples incorrectly flagged.
The per-class metrics are
\begin{align}
  \text{Accuracy}  &= \frac{TP + TN}{TP + TN + FP + FN}, \label{eq:acc}\\[3pt]
  \text{Precision} &= \frac{TP}{TP + FP}, \qquad
  \text{Recall}    = \frac{TP}{TP + FN}, \label{eq:prec_rec}\\[3pt]
  \text{F1}        &= 2 \cdot \frac{\text{Precision} \cdot \text{Recall}}
                                    {\text{Precision} + \text{Recall}}. \label{eq:f1}
\end{align}
For the multiclass Stage~2 classifier, the macro-averaged F1-score assigns equal weight to every
fault class regardless of its support, which is necessary under the severe class imbalance of the
primary dataset,
\begin{equation}
  \text{macro-F1} = \frac{1}{|\mathcal{C}|} \sum_{c \in \mathcal{C}} \text{F1}_c,
  \label{eq:macrof1}
\end{equation}
where $\text{F1}_c$ is the F1-score of class $c$.

Two distinct accuracy figures are reported and should not be conflated. \emph{Stage~2 accuracy} is
measured over true fault samples that Stage~1 has already detected, and therefore excludes both
normal samples and faults missed by Stage~1; it characterizes the classifier in isolation.
\emph{End-to-end accuracy} is measured over the complete test set, including normal samples and
counting Stage~1 misses as errors,
\begin{equation}
  \text{Acc}_{\text{e2e}} = \frac{1}{N}\sum_{i=1}^{N}
    \mathbb{1}\!\left[\hat{y}_{\text{e2e}}(x_i) = y_i\right],
  \qquad
  \hat{y}_{\text{e2e}}(x) =
  \begin{cases}
    \hat{y}(x) & \text{if } d_{\text{S1}}(x) = 1,\\
    \text{normal} & \text{otherwise,}
  \end{cases}
  \label{eq:e2e}
\end{equation}
and is the quantity comparable to the accuracy reported by TLFed~\cite{custodio2023}. Per-fault-type
end-to-end accuracy is the product of the Stage~1 detection rate and the Stage~2 classification
accuracy for that type, since a sample is handled correctly only if it is both detected and then
correctly classified.

% ============================================================
\section{Datasets}
% ============================================================

The framework is evaluated on two independent datasets, summarized in Table~\ref{tab:datasets}. They
differ in network size, voltage and current scale, simulation software, class balance, and label
encoding, and were produced by different authors. The first serves as the primary evaluation system
and supports comparison against the TLFed benchmark; the second provides independent validation of
the architecture and of the feature construction.

\begin{table}[H]
  \centering
  \caption{Summary of the two evaluation datasets.}
  \label{tab:datasets}
  \begin{tabular}{lcc}
    \toprule
    \textbf{Property} & \textbf{TLFaultDataset} & \textbf{Second dataset} \\
    \midrule
    Measurement points $L$    & 7         & 1 \\
    Raw channels              & 42        & 6 \\
    Engineered features       & 126       & 18 (16 informative) \\
    Samples                   & 578{,}923 & 12{,}001 \\
    Normal proportion         & 88.1\%    & 54.2\% \\
    Fault classes             & 6         & 5 \\
    Near-normal class present & Yes (Line fault) & No \\
    Fault location labeled    & Yes       & No \\
    \bottomrule
  \end{tabular}
\end{table}

\subsection{Primary Dataset: TLFaultDataset}

The TLFaultDataset is a simulated power transmission network dataset containing 578{,}923 time-series
samples from a network with seven transmission lines. Each line is named by the pair of buses it
connects: L12 denotes the line between bus~1 and bus~2, L13 the line between bus~1 and bus~3, and so
on for L23, L24, L25, L34, and L45. These line identifiers refer to physical network elements and
should not be confused with the fault-type labels defined below. For every sample, three-phase
currents ($I_a, I_b, I_c$) and three-phase voltages ($V_a, V_b, V_c$) are recorded for each of the
seven lines, yielding $7 \times 6 = 42$ raw features. Each sample is labeled with one of seven fault
categories: No fault (normal), DLG (double line-to-ground), L-L (line-to-line), SLG (single
line-to-ground), Line fault, L-L-L (three-phase), and L-L-L-G (three-phase-to-ground). The
\texttt{Number} column serves as a time index, and samples are ordered by it to preserve temporal
ordering.

The class distribution is severely imbalanced, as shown in Table~\ref{tab:dist}. Normal operation
constitutes 88.1\% of all samples, while the rarest fault (L-L-L-G) accounts for under 1\%. Line
faults, despite being the most difficult to detect in practice, represent only 1.4\% of the dataset.
A critical characteristic of Line faults is that their three-phase measurements are nearly identical
to normal operation, producing only subtle changes in the current of the affected line and rendering
them invisible to threshold-based methods and to unsupervised anomaly detectors trained on normal
data.

\begin{table}[H]
  \centering
  \caption{Fault type distribution in the TLFaultDataset.}
  \label{tab:dist}
  \begin{tabular}{llrr}
    \toprule
    \textbf{Fault Type} & \textbf{Description} & \textbf{Samples} & \textbf{Percentage} \\
    \midrule
    No fault   & Normal operation              & 509{,}754 & 88.1\% \\
    DLG        & Double line-to-ground fault   &  17{,}682 &  3.1\% \\
    L-L        & Line-to-line fault            &  17{,}095 &  3.0\% \\
    SLG        & Single line-to-ground fault   &  13{,}673 &  2.4\% \\
    Line fault & Transmission line fault       &   7{,}875 &  1.4\% \\
    L-L-L      & Three-phase fault             &   7{,}254 &  1.3\% \\
    L-L-L-G    & Three-phase-to-ground fault   &   5{,}590 &  1.0\% \\
    \midrule
    \textbf{Total} &                           & \textbf{578{,}923} & \textbf{100\%} \\
    \bottomrule
  \end{tabular}
\end{table}

\subsection{Second Dataset: Independent Validation}
\label{sec:dataset2}

The second dataset is a publicly available simulated fault dataset for a single-measurement-point
transmission system~\cite{kaggledataset}. It was produced independently of the TLFaultDataset, using
different simulation software and a different network, with per-unit voltages and phase currents in
the range $\pm 900$~A. It is used here to determine whether the architecture and the feature
construction transfer to a system on which the method was not developed.

\textbf{Composition.} The dataset is distributed as two files. Verification established that all
7{,}861 rows of the multiclass file are contained within the 12{,}001-row detection file, which
additionally contains 4{,}140 normal samples; the files are therefore not independent partitions and
were merged into a single table before splitting, since treating them as separate sources would place
identical samples on both sides of the train/test split. The merged dataset contains 6{,}505 normal
and 5{,}496 fault samples, a fault ratio of 45.8\%. Class counts are given in Table~\ref{tab:dist2}.

\begin{table}[H]
  \centering
  \caption{Class distribution in the second dataset after merging.}
  \label{tab:dist2}
  \begin{tabular}{lrr}
    \toprule
    \textbf{Fault Type} & \textbf{Samples} & \textbf{Percentage} \\
    \midrule
    No fault   & 6{,}505 & 54.2\% \\
    DLG        & 1{,}134 &  9.4\% \\
    L-L-L-G    & 1{,}133 &  9.4\% \\
    SLG        & 1{,}129 &  9.4\% \\
    L-L-L      & 1{,}096 &  9.1\% \\
    L-L        & 1{,}004 &  8.4\% \\
    \midrule
    \textbf{Total} & \textbf{12{,}001} & \textbf{100\%} \\
    \bottomrule
  \end{tabular}
\end{table}

\textbf{Label derivation.} The dataset encodes ground truth as four binary indicator columns
recording which phases are involved and whether the fault involves ground, rather than as fault-type
names. These indicators were mapped to the standard symmetrical-fault taxonomy by phase count and
ground involvement: one phase with ground gives SLG; two phases without ground gives L-L and with
ground gives DLG; three phases without ground gives L-L-L and with ground gives L-L-L-G. This is a
re-encoding of the dataset's own ground truth into the vocabulary used by the TLFaultDataset, and it
is what permits the per-class comparison of Section~\ref{sec:crossdataset}. The taxonomy admits a
single-phase open-conductor category that is not populated in this dataset.

\textbf{Differences from the primary dataset.} Three differences are relevant to interpretation. The
class distribution is close to balanced, so the imbalance motivation of Section~1 does not apply; the
dataset contains no near-normal class analogous to Line fault; and it provides no fault-location
label, so the localization extension discussed in Section~\ref{sec:discussion} cannot be evaluated on
it.

\subsection{Preprocessing}

Both datasets pass through the same preprocessing sequence. Where a time index is available, data is
sorted by it. A stratified 80/20 train/test split on the binary fault label preserves class
proportions. The standardization transform is fitted exclusively on normal training samples. A
validation split is held out from the training data for Stage~1 threshold selection and for the
Branch~B allocation of~\eqref{eq:allocation}; the test set remains untouched until final evaluation.
All feature arrays are stored in \texttt{float32} precision, halving memory consumption without
measurable effect on model quality. For the TLFaultDataset this yields 463{,}135 training and
115{,}784 test samples, with the transform fitted on 407{,}800 normal training samples; for the
second dataset, 9{,}600 training and 2{,}401 test samples.

% ============================================================
\section{Results}
% ============================================================

\subsection{Stage 1: Hybrid Fault Detection (Primary Dataset)}

Table~\ref{tab:stage1} compares the Isolation Forest alone against the fused Stage~1 detector on the
TLFaultDataset. The allocation rule of~\eqref{eq:allocation} at $\tau = 0.5$ selects the Line-fault
class, whose Branch~A detection rate of 0.9\% is the only rate below threshold. The fused scheme
improves detection across every fault type, with the largest gain for Line faults, where the
dedicated detector raises detection from 0.9\% to \needsupdate{98.5\%}. This result supports the core
architectural decision: the Isolation Forest is structurally blind to faults whose signatures overlap
with normal operation, and only a supervised detector trained with explicit labels for that class
addresses the limitation. The trade-off is an increased false alarm rate, which is acceptable in a
safety-critical setting. Raising the Branch~B decision threshold from 0.50 to 0.70 reduces false
alarms at the cost of Line-fault recall.

\begin{table}[H]
  \centering
  \caption{Stage 1 detection on the TLFaultDataset: Isolation Forest alone versus the fused detector.}
  \label{tab:stage1}
  \begin{tabular}{lcc}
    \toprule
    \textbf{Fault Type / Metric} & \textbf{Branch A alone} & \textbf{Stage 1 fused} \\
    \midrule
    Line fault detection & 0.9\%  & \needsupdate{98.5\%} \\
    L-L-L detection      & \needsupdate{95.2\%} & \needsupdate{95.8\%} \\
    L-L-L-G detection    & \needsupdate{93.1\%} & \needsupdate{94.6\%} \\
    DLG detection        & \needsupdate{93.2\%} & \needsupdate{94.5\%} \\
    L-L detection        & \needsupdate{91.8\%} & \needsupdate{93.9\%} \\
    SLG detection        & \needsupdate{87.9\%} & \needsupdate{88.9\%} \\
    \midrule
    Fault recall         & \needsupdate{81.5\%} & \needsupdate{92.0\%} \\
    False alarm rate     & \needsupdate{1.5\%}  & \needsupdate{5.3\%}  \\
    \bottomrule
  \end{tabular}
\end{table}

\subsection{Impact of Zero-Sequence Features (Primary Dataset)}

\begin{figure}[H]
  \centering
  \includegraphics[width=0.85\linewidth]{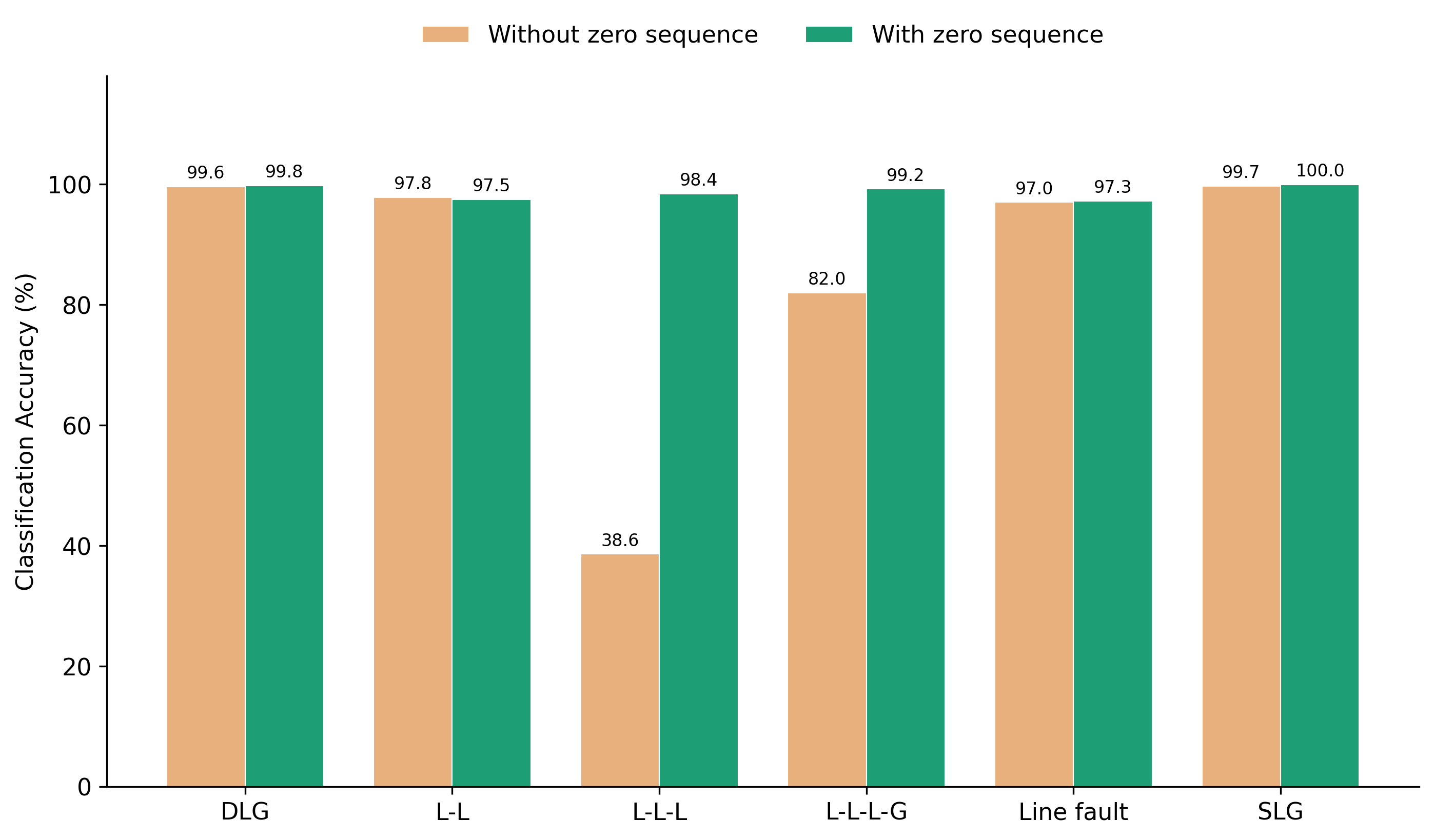}
  \caption{Impact of zero-sequence features on Stage 2 classification accuracy (TLFaultDataset).}
  \label{fig:zeroseq}
\end{figure}

Fig.~\ref{fig:zeroseq} reports Stage~2 classification accuracy before and after adding zero-sequence
features on the TLFaultDataset. Without them, the classifier confuses L-L-L and L-L-L-G in the
majority of cases, because their raw phase voltages and currents are nearly identical in steady
state. Adding the zero-sequence group raises L-L-L accuracy from 38.6\% to \needsupdate{98.4\%} and
L-L-L-G from 82.0\% to \needsupdate{99.2\%}.

\subsection{Stage 2: Fault Type Classification (Primary Dataset)}

\begin{figure}[H]
  \centering
  \includegraphics[width=0.7\linewidth]{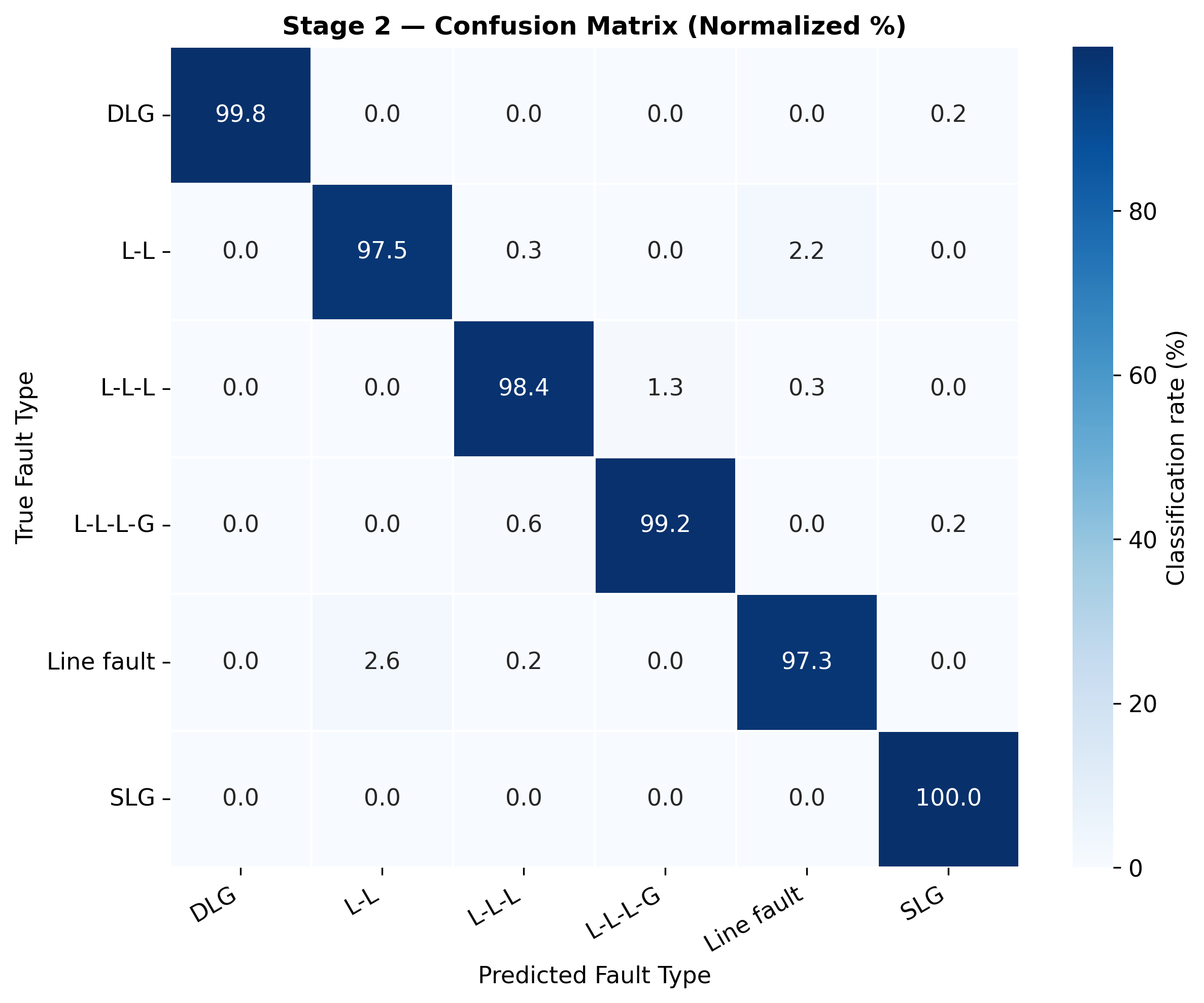}
  \caption{Stage 2 normalized confusion matrix, TLFaultDataset.}
  \label{fig:cm}
\end{figure}

Table~\ref{tab:stage2} reports the Stage~2 classification report on true fault samples detected by
Stage~1. Stage~2 achieves \needsupdate{99\%} accuracy with a macro F1-score of \needsupdate{0.99} on
this filtered input. DLG and SLG achieve the highest F1-scores, consistent with their distinctive
signatures. L-L-L and L-L-L-G each reach F1 of \needsupdate{0.99}, resolving the confusion that
reduced L-L-L accuracy to 38.6\% without zero-sequence features. Fig.~\ref{fig:cm} shows the
corresponding normalized confusion matrix.

\begin{table}[H]
  \centering
  \caption{Stage 2 classification report on true fault samples detected by Stage 1 (TLFaultDataset).}
  \label{tab:stage2}
  \begin{tabular}{lcccc}
    \toprule
    \textbf{Fault Type} & \textbf{Precision} & \textbf{Recall} & \textbf{F1-Score} & \textbf{Support} \\
    \midrule
    DLG        & \needsupdate{1.00} & \needsupdate{1.00} & \needsupdate{1.00} & \needsupdate{3{,}474} \\
    L-L        & \needsupdate{0.99} & \needsupdate{0.97} & \needsupdate{0.98} & \needsupdate{3{,}236} \\
    L-L-L      & \needsupdate{0.99} & \needsupdate{0.98} & \needsupdate{0.99} & \needsupdate{1{,}409} \\
    L-L-L-G    & \needsupdate{0.98} & \needsupdate{0.99} & \needsupdate{0.99} & \needsupdate{1{,}026} \\
    Line fault & \needsupdate{0.95} & \needsupdate{0.97} & \needsupdate{0.96} & \needsupdate{1{,}565} \\
    SLG        & \needsupdate{1.00} & \needsupdate{1.00} & \needsupdate{1.00} & \needsupdate{2{,}544} \\
    \midrule
    \textbf{Macro avg} & \needsupdate{\textbf{0.98}} & \needsupdate{\textbf{0.99}} &
                         \needsupdate{\textbf{0.99}} & \needsupdate{\textbf{13{,}254}} \\
    \bottomrule
  \end{tabular}
\end{table}

\subsection{End-to-End Performance (Primary Dataset)}

\begin{figure}[H]
  \centering
  \includegraphics[width=0.85\linewidth]{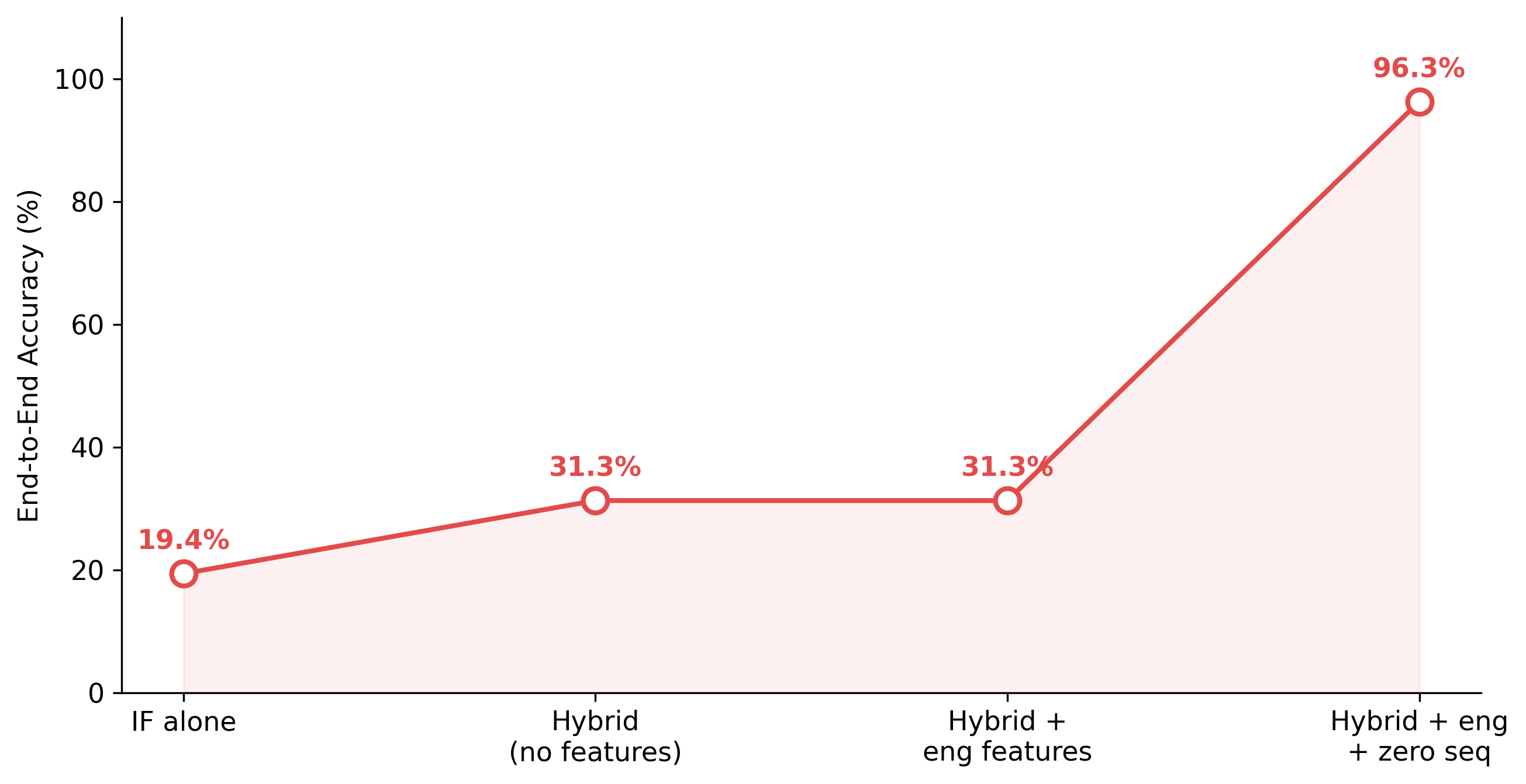}
  \caption{Line-fault end-to-end accuracy across methodology iterations.}
  \label{fig:journey}
\end{figure}

Table~\ref{tab:e2e} combines Stage~1 detection and Stage~2 classification into end-to-end accuracy
per fault type on the TLFaultDataset. All six fault types exceed \needsupdate{88\%} end-to-end
accuracy. Fig.~\ref{fig:journey} traces the Line-fault improvement across methodology iterations: the
Isolation Forest alone achieves 19.4\% end-to-end accuracy for Line faults; adding the dedicated
detector raises this to 31.3\%, a modest gain because Stage~2 still receives few Line-fault samples;
and the complete pipeline with zero-sequence features reaches \needsupdate{95.8\%}.

\begin{table}[H]
  \centering
  \caption{End-to-end performance by fault type, TLFaultDataset.}
  \label{tab:e2e}
  \begin{tabular}{lccc}
    \toprule
    \textbf{Fault Type} & \textbf{Stage 1 Detection} & \textbf{Stage 2 Classification} & \textbf{End-to-End} \\
    \midrule
    Line fault & \needsupdate{98.5\%} & \needsupdate{97.3\%} & \needsupdate{95.8\%} \\
    DLG        & \needsupdate{94.5\%} & \needsupdate{99.8\%} & \needsupdate{94.3\%} \\
    L-L-L      & \needsupdate{95.8\%} & \needsupdate{98.4\%} & \needsupdate{94.3\%} \\
    L-L-L-G    & \needsupdate{94.6\%} & \needsupdate{99.2\%} & \needsupdate{93.8\%} \\
    L-L        & \needsupdate{93.9\%} & \needsupdate{97.5\%} & \needsupdate{91.5\%} \\
    SLG        & \needsupdate{88.9\%} & \needsupdate{99.9\%} & \needsupdate{88.8\%} \\
    \bottomrule
  \end{tabular}
\end{table}

% ------------------------------------------------------------
\subsection{Validation on the Second Dataset}
\label{sec:results2}
% ------------------------------------------------------------

The framework was applied to the second dataset without modification to the architecture. The only
dataset-specific component is the input adapter that declares which columns constitute the
three-phase channel groups; the feature operator $\varphi$, the allocation rule, the fusion rule, and
both classifiers are unchanged. With $L = 1$, \eqref{eq:featcount} gives 18 features, of which 16 are
informative.

\begin{figure}[H]
  \centering
  \begin{subfigure}[b]{0.48\linewidth}
    \includegraphics[width=\linewidth]{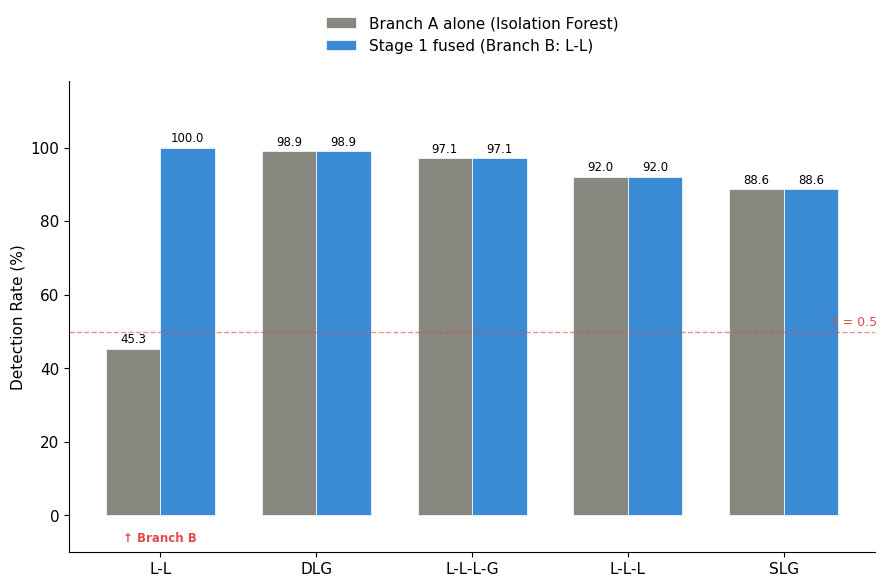}
    \caption{Stage 1 detection by fault type.}
  \end{subfigure}
  \hfill
  \begin{subfigure}[b]{0.48\linewidth}
    \includegraphics[width=\linewidth]{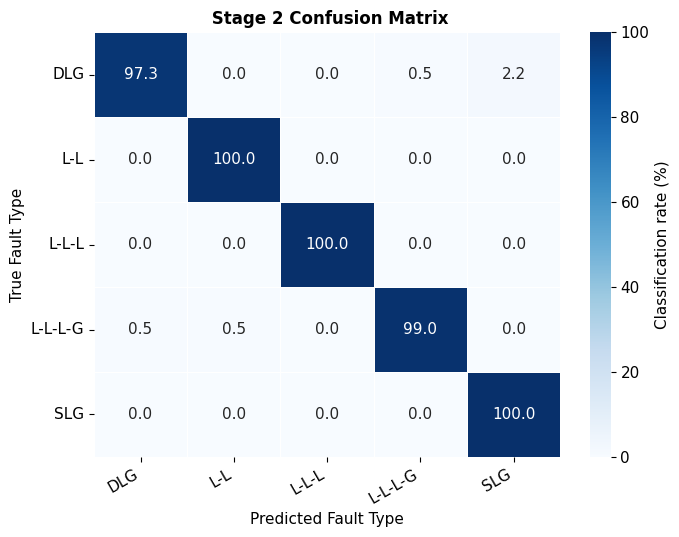}
    \caption{Stage 2 normalized confusion matrix.}
  \end{subfigure}
  \caption{Second-dataset Stage 1 and Stage 2 results. In (a) the dashed line marks the allocation
           threshold $\tau = 0.5$; only L-L falls below it and receives a Branch~B detector, while the
           remaining classes are unchanged by fusion.}
  \label{fig:stage1_2}
\end{figure}

\textbf{Branch B allocation.} On the validation split the Isolation Forest detected the L-L class at
45.3\%, below $\tau = 0.5$, while all other classes exceeded 88\%. The allocation rule therefore
instantiated a single supervised detector for L-L, raising its detection to 100\% at a false alarm
rate of 0.31\%. The rule selected a different class than on the primary dataset, where it selects
Line fault, indicating that the mechanism responds to the properties of the system rather than to a
predetermined class identity. Fig.~\ref{fig:stage1_2}(a) shows the detection rates before and after
fusion, with the allocation threshold marked.

\textbf{Detection and classification.} Stage~1 achieved 95.00\% fault recall at a 0.31\% false alarm
rate. Stage~2 achieved 99.33\% accuracy on detected faults. End-to-end accuracy computed
by~\eqref{eq:e2e} over all 2{,}401 test samples, including normal operation, is \textbf{97.25\%},
exceeding the TLFed benchmark of 94.84\%. Per-class results are given in Table~\ref{tab:e2e2} and the
confusion matrix in Fig.~\ref{fig:stage1_2}(b).

\begin{table}[H]
  \centering
  \caption{End-to-end performance by fault type, second dataset.}
  \label{tab:e2e2}
  \begin{tabular}{lccc}
    \toprule
    \textbf{Fault Type} & \textbf{Stage 1 Detection} & \textbf{Stage 2 Classification} & \textbf{End-to-End} \\
    \midrule
    L-L      & 100.0\% & 100.0\% & 100.0\% \\
    DLG      &  98.9\% &  97.3\% &  96.3\% \\
    L-L-L-G  &  97.1\% &  99.0\% &  96.1\% \\
    L-L-L    &  92.0\% & 100.0\% &  92.0\% \\
    SLG      &  88.6\% & 100.0\% &  88.6\% \\
    \bottomrule
  \end{tabular}
\end{table}

\begin{figure}[H]
  \centering
  \includegraphics[width=0.92\linewidth]{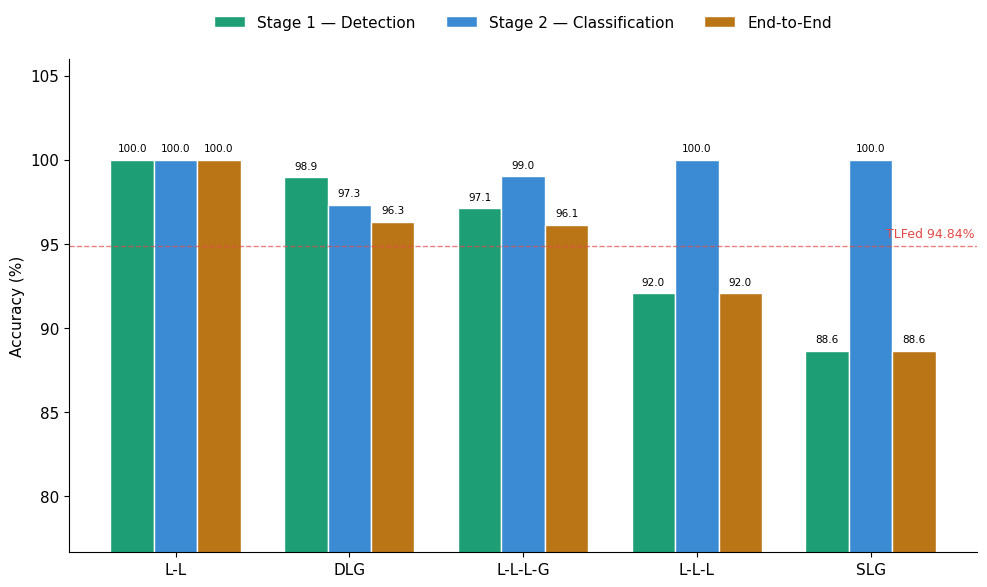}
  \caption{Per-class Stage 1, Stage 2, and end-to-end accuracy on the second dataset. The end-to-end
           value is the product of the two stage-wise values. The dashed line marks the TLFed
           benchmark; the comparable aggregate is the overall end-to-end accuracy of 97.25\%.}
  \label{fig:e2e2fig}
\end{figure}

\textbf{Distribution of error across stages.} Fig.~\ref{fig:e2e2fig} decomposes performance into its
two stages for each fault class. The end-to-end bar is the product of the two preceding bars, and in
every class it is the smallest of the three, indicating that detection rather than classification
governs overall performance on this dataset. For SLG, Stage~2 classifies every detected sample
correctly, so the 88.6\% end-to-end figure is attributable entirely to Stage~1 misses; the same
pattern holds for L-L-L at 92.0\%. DLG is the only class for which Stage~2 contributes measurable
error, at 97.3\%, corresponding to the 2.2\% of DLG samples assigned to SLG in
Fig.~\ref{fig:stage1_2}(b)---a confusion between two fault types that differ by a single involved
phase. The practical implication is that further improvement on this dataset would come from
raising Stage~1 recall rather than from a stronger classifier. The horizontal reference line marks
the TLFed benchmark of 94.84\%; the comparable quantity is the overall end-to-end accuracy of
97.25\% rather than the individual per-class bars, since the benchmark is itself an aggregate figure.

% ------------------------------------------------------------
\subsection{Cross-Dataset Feature Ablation}
\label{sec:crossdataset}
% ------------------------------------------------------------

\begin{figure}[H]
  \centering
  \includegraphics[width=0.92\linewidth]{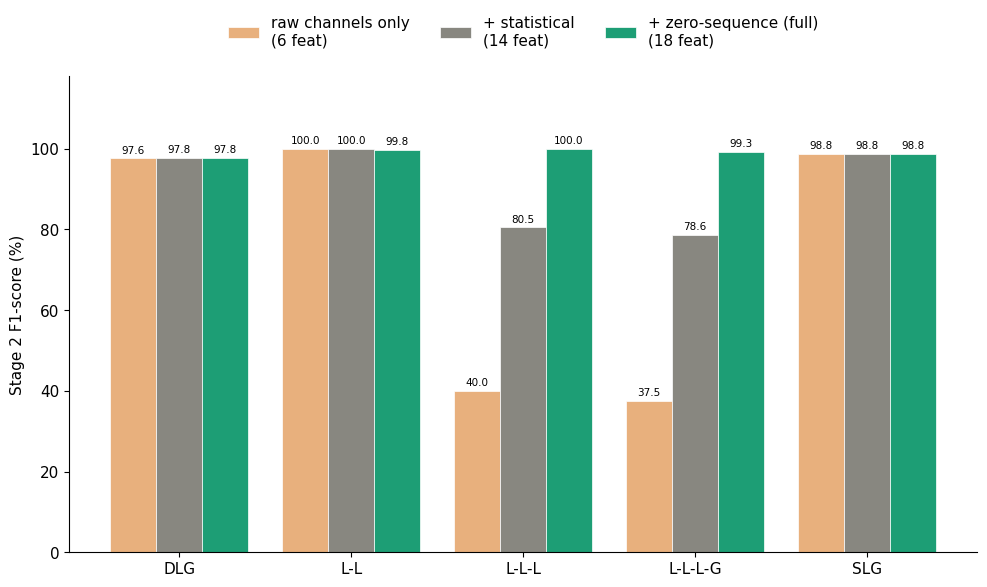}
  \caption{Stage 2 feature ablation on the second dataset. DLG, L-L, and SLG are resolved from the
           raw channels alone and are unaffected by the added groups; the entire benefit of the
           zero-sequence features accrues to the L-L-L and L-L-L-G pair.}
  \label{fig:ablation2fig}
\end{figure}

Table~\ref{tab:ablation2} and Fig.~\ref{fig:ablation2fig} report a Stage~2 feature ablation on the
second dataset, isolating the contribution of each feature group. From the six raw channels alone the
classifier cannot separate L-L-L from L-L-L-G, achieving F1-scores of 0.400 and 0.375
respectively close to the value expected from assigning the pair at random while the three
remaining classes are already resolved above 0.97. Adding the statistical descriptors lifts the pair
to approximately 0.80. Adding the zero-sequence group resolves it, at 1.000 and 0.993.

\begin{table}[H]
  \centering
  \caption{Stage 2 feature ablation on the second dataset (F1-score per class).}
  \label{tab:ablation2}
  \begin{tabular}{lcccccc}
    \toprule
    \textbf{Feature set} & \textbf{\# Feat.} & \textbf{Macro F1} &
    \textbf{L-L-L} & \textbf{L-L-L-G} & \textbf{DLG} & \textbf{SLG} \\
    \midrule
    Raw channels only     &  6 & 0.748 & 0.400 & 0.375 & 0.976 & 0.988 \\
    $+$ statistical       & 14 & 0.912 & 0.805 & 0.786 & 0.978 & 0.988 \\
    $+$ zero-sequence     & 18 & \textbf{0.991} & \textbf{1.000} & \textbf{0.993} & 0.978 & 0.988 \\
    \bottomrule
  \end{tabular}
\end{table}

The selectivity visible in Fig.~\ref{fig:ablation2fig} is as informative as the magnitude of the
improvement. DLG, L-L, and SLG vary by less than one percentage point across all three feature sets,
so the engineered features neither help nor harm classes that are already separable from the raw
measurements. The gain is confined to the single class pair for which the underlying physical
argument predicts it, which is consistent with the zero-sequence group encoding ground involvement
specifically rather than acting as a generic increase in representational capacity.

This reproduces, on an independent system, the effect reported for the TLFaultDataset in
Fig.~\ref{fig:zeroseq}: the discriminative information separating three-phase from
three-phase-to-ground faults is absent from the raw measurements and is supplied by the zero-sequence
components. That the same feature group resolves the same class pair on two systems differing in
size, scale, and simulation software indicates that the effect follows from the physics rather than
from a property of one dataset.

% ------------------------------------------------------------
\subsection{System Dependence of the Zero-Sequence Signature}
\label{sec:zerosysdep}
% ------------------------------------------------------------

While the zero-sequence \emph{features} resolve the three-phase class pair on both systems, the
\emph{direction} of the signature is not consistent between them. On the TLFaultDataset, L-L-L-G
samples exhibit larger $|I_0|$ than L-L-L samples, consistent with the expectation that a ground path
admits residual current. On the second dataset the ordering is reversed: the median $|I_0|$ for
L-L-L-G is $1 \times 10^{-3}$ against $0.70$ for L-L-L, and ranking samples by $|I_0|$ is
anti-predictive for the pair. A fixed threshold on $|I_0|$ calibrated on one system would therefore
misclassify the pair on the other.

The likely physical explanation is that zero-sequence current is a residual quantity requiring both a
ground path and an asymmetry among the phase currents. A balanced bolted three-phase-to-ground fault,
in which the three phase impedances to ground are equal, leaves the phase currents symmetric; they
continue to sum to approximately zero and little residual flows despite the ground path being
present. The three-phase-to-ground events in the second dataset appear to be predominantly of this
balanced type, whereas those in the TLFaultDataset carry sufficient asymmetry---arising from unequal
fault resistances, non-simultaneous pole closure, or unbalanced source impedance---to produce a
measurable residual. The 99th percentile of $|I_0|$ for L-L-L-G on the second dataset is
$2.6 \times 10^{2}$, confirming that a minority of asymmetric events do produce a large residual. A
controlled parameter sweep over fault resistance asymmetry would be required to establish this
mechanism directly and is identified as future work in Section~\ref{sec:discussion}.

This observation has a direct methodological consequence. The magnitude and sign of the zero-sequence
signature depend on fault balance and grounding impedance, which are properties of the installation
rather than of the fault type. A fixed-threshold relay rule based on $|I_0|$ is therefore not portable
across systems, whereas a decision boundary learned from the zero-sequence feature group together
with the remaining features attains F1 above 0.99 for the pair on both systems. This constitutes an
argument for a learned classifier over a threshold rule in this setting, and it is an argument that
can be made only once more than one system has been examined.

% ------------------------------------------------------------
\subsection{Computational Cost}
\label{sec:cost}
% ------------------------------------------------------------

Inference cost was measured on CPU without GPU acceleration. On the second dataset the complete
pipeline Branch~A scoring, Branch~B scoring, and Stage~2 classification requires 0.052~ms per
sample, corresponding to a throughput of approximately $1.9 \times 10^{4}$ samples per second on a
single machine. The comparison against TLFed is therefore not restricted to accuracy: the present
pipeline exceeds the federated 1D-CNN-LSTM benchmark while requiring neither GPU hardware nor a
distributed training infrastructure, which is relevant for deployment in substation environments
where computational resources are constrained.

% ============================================================
\section{Discussion and Future Work}
\label{sec:discussion}
% ============================================================

The experimental results support three principal observations.

First, the improvement in Line-fault performance is attributable to the architecture rather than to
the classifier. In the feature space provided to the Isolation Forest, Line-fault samples are not
statistically distinguishable from normal operating points, and neither increasing the number of
trees nor retuning the contamination rate addresses this limitation. The improvement instead results
from introducing a dedicated supervised detector targeting the specific class that overlaps with
normal operation. The allocation rule of~\eqref{eq:allocation} generalizes this observation: it
identifies the Line-fault class on the primary dataset and the L-L class on the second, without
either being specified in advance. The two cases differ in degree, however. Line fault is detected at
0.9\% on the primary dataset and remains below the allocation threshold at every Stage~1 operating
point examined, whereas L-L on the second dataset is detected at 45.3\% at the reported operating
point and rises above the threshold at stricter recall targets. The rule identifies a structurally
invisible class in the first case and a marginally weak one in the second, and this distinction
should be borne in mind when interpreting the generality of the mechanism.

Second, the zero-sequence result demonstrates the same principle from the feature perspective, and
the cross-dataset evidence strengthens rather than merely confirms it. Without $I_0$ and $V_0$, the
classifier must separate L-L-L from L-L-L-G using signals that are nearly identical in steady state,
and the majority of samples in the pair are misassigned; this holds on both systems. Introducing the
zero-sequence components resolves the ambiguity on both. As Section~\ref{sec:zerosysdep} establishes,
however, the direction of the signature differs between the two systems, so the resolution cannot be
attributed to a fixed physical rule of the form ``$|I_0|$ is larger for ground faults.'' The
discriminative information is present in the zero-sequence subspace, but the boundary separating the
classes is system-dependent and must be learned. For this class of problem, physically grounded
feature engineering combined with a learned boundary appears to contribute more than either increased
model complexity or a fixed threshold rule.

Third, the framework transfers across systems while the fitted models do not. Retraining the same
architecture, feature operator, and allocation rule on each dataset yields strong performance on
both. A model trained on one dataset and applied directly to the other, however, performs near
chance, with the dominant failure mode being the inversion of the three-phase class pair described in
Section~\ref{sec:zerosysdep}. This is consistent with established protection practice, in which relay
settings are commissioned per installation rather than transferred between substations, and it
reinforces the argument for a learned per-system boundary.

The principal cost of the design is an increased false alarm rate on the primary dataset, which rises
with the addition of the Branch~B detector. In a protection context this trade-off is acceptable,
since a missed fault carries substantially greater consequences than an investigated false alarm;
in high-volume monitoring, however, the additional alarms impose an operational burden. The decision
threshold of the Branch~B detector provides direct control over this trade-off, and the appropriate
operating point is determined by the priorities of the target deployment.

Several limitations remain and define directions for future work. Both datasets are simulated and
therefore do not reproduce the measurement noise, sensor drift, harmonics, and transient disturbances
present in operational substation measurements; validation on field data remains the primary next
step, and a controlled noise-injection study is a necessary intermediate. Establishing the mechanism
proposed in Section~\ref{sec:zerosysdep} requires a parameter sweep over fault resistance asymmetry
and grounding impedance, which a simulation environment would permit under controlled conditions. The
pipeline processes each time step independently and incorporates no temporal context, so slowly
developing faults are not informed by their own history; a sliding-window or recurrent front end is a
candidate extension, although preliminary experiments with a sliding-window input did not yield
improvements sufficient to justify the additional cost. Finally, the pipeline identifies the fault
type but not its location along the line, which is equally important for repair dispatch. Fault
localization is a natural third stage that would operate on the same detected-fault samples already
provided to Stage~2, and would permit comparison against the 92.50\% location accuracy reported by
TLFed; the primary dataset carries the requisite labels. Extension to microgrids with inverter-based
resources, in which fault-current behavior differs substantially from conventional transmission,
is a longer-term objective.

% ============================================================
\section{Conclusion}
% ============================================================

This paper addressed two challenges that limit single-model fault classifiers: datasets dominated by
normal samples, and fault types whose electrical signatures are nearly indistinguishable from normal
operation. Decomposing the task into a detection stage optimized for recall and a classification
stage optimized for per-class precision addresses both. Detection combines an unsupervised Isolation
Forest with a supervised binary detector allocated automatically for any class the anomaly detector
cannot resolve; on the primary dataset this raised Line-fault end-to-end accuracy from 31.3\% to
\needsupdate{95.8\%}.

The framework was evaluated on two independent datasets differing in network size, scale, simulation
software, and class balance, with the feature construction expressed as a per-measurement-point
operator applied identically at $L = 7$ and $L = 1$. On the second dataset the pipeline attained
97.25\% end-to-end accuracy across all classes including normal operation, exceeding the TLFed
benchmark of 94.84\% without federated or GPU infrastructure and at 0.05~ms per sample on CPU. The
allocation rule identified a blind-spot class on both systems without either being specified in
advance, and ablation showed the zero-sequence feature group resolving the three-phase versus
three-phase-to-ground ambiguity on both, raising the F1-score of that pair from 0.39 to 0.997 on the
second dataset while leaving the remaining classes unchanged.

The cross-dataset evaluation also produced a result not visible from a single system: the direction
of the zero-sequence signature depends on fault balance and grounding impedance and is not consistent
between the two networks, so a fixed threshold calibrated on one would fail on the other while a
learned boundary succeeds on both. These results indicate that, for fault types that overlap with
normal operation or with one another, domain-informed feature engineering combined with a
problem-specific architecture and a learned decision boundary provides greater benefit than increased
model complexity.

% ============================================================
%  References
% ============================================================

\end{document}